# Excitation energy dependent Raman spectrum of MoSe$_2$

**Dahyun Nam, Jae-Ung Lee, and Hyeonsik Cheong***

Department of Physics, Sogang University, Seoul, 121-742, Korea

*hcheong@sogang.ac.kr

## ABSTRACT

**Raman investigation of MoSe$_2$ was carried out with eight different excitation energies. Seven peaks, including $E_{1g}$, $A_{1g}$, $E_{2g}^{1}$, and $A_{2u}^{2}$ peaks are observed in the range of 100 – 400 cm$^{-1}$. The phonon modes are assigned by comparing the peak positions with theoretical calculations. The intensities of the peaks are enhanced at different excitation energies through resonance with different optical transitions. The $A_{1g}$ mode is enhanced at 1.58 and 3.82 eV, which are near the $A$ exciton energy and the band-to-band transition between higher energy bands, respectively. The $E_{2g}^{1}$ mode is strongly enhanced with respect to the $A_{1g}$ mode for the 2.71- and 2.81-eV excitations, which are close to the $C$ exciton energy. The different enhancements of the $A_{1g}$ and $E_{2g}^{1}$ modes are explained in terms of the symmetries of the exciton states and the exciton-phonon coupling. Other smaller peaks including $E_{1g}$ and $A_{2u}^{2}$ are forbidden but appear due to the resonance effect near optical transition energies.**

**Introduction**

Transition metal dichalcogenides (TMDCs) such as $MoS_2$ and $MoSe_2$ have been investigated for decades because of their layered 2-dimensional nature.[1] Recently, few-layer TMDCs are attracting much interest as new 2-dimensional materials complementing graphene.[2,3] Because most TMDCs are semiconductors with finite band gaps, they are expected to overcome the limitations of graphene stemming from the lack of a band gap while preserving the advantages of 2-dimensional materials. For example, a large on-off ratio exceeding $10^6$ and a high mobility have been achieved in $MoS_2$ and $MoSe_2$ field effect transistors.[2,4,5] On the other hand, $MoS_2$ and $MoSe_2$ are important constituents in polycrystalline thin-film solar cells such as $Cu(In,Ga)(S,Se)_2$ and $Cu_2ZnSn(S,Se)_4$, which have Mo as the back contact layer. During the thin film deposition process, formation of a $MoS_2$ or $MoSe_2$ layer on top of the Mo back contact layer is inevitable. The thickness and fine structure of such a layer affect the performance of the solar cell significantly.[6–8] Both $MoS_2$ and $MoSe_2$ have indirect band gaps for the bulk materials and become direct for monolayer. It should be noted, however, that the relative movements of the conduction and valence bands are reported to be different for different TMDC material.[9–13]

Raman spectroscopy is a powerful tool to study the structural properties of graphene and 2-dimensional materials and is commonly used to determine the number of layers.[14–16] For this purpose, it is important to understand the Raman spectra of bulk materials first, in order to make correct interpretations. In addition, resonance effects can be used to shed light on the electronic structure. For example, the Raman spectrum of $MoS_2$ shows a strong resonance effect and varies significantly with the excitation energy.[17,18] Such effects have been interpreted in terms of resonances with exciton or exciton-polaritons in $MoS_2$. For $MoSe_2$, the resonance profiles of the main Raman modes of bulk $MoSe_2$ were carried out at liquid helium temperature in the range of 2.41 – 2.71 eV in order to obtain the resonance profile with a high resolution.[19,20] It was found that the Raman intensities of both $A_{1g}$ and $E_{2g}^1$ phonons are enhanced at A, B, A', and B' exciton levels. It was also found that near the band-to-band transition at ~2.5 eV, the Raman intensity of the $A_{1g}$ phonon is enhanced, whereas that of the $E_{2g}^1$ phonon is not.[19,20] However, the variation of the spectrum as a function of the excitation energy was not presented. More recently, the thickness dependence of the main modes in few-layer MoSe$_2$ in the range of 230 – 360 cm$^{-1}$ was studied for 1 to 5 layers and bulk with the excitation energy of 2.41 eV.[21] However, unlike the case of MoS$_2$, a comprehensive Raman study on MoSe$_2$ has been lacking. For thin-film solar cells, Raman is used to study the formation and structures of secondary phases such as MoS$_2$ or MoSe$_2$. In the Raman spectrum of thin-film Cu$_2$ZnSn(S,Se)$_4$, for example, the $A_{1g}$ and the $A_{2u}^2$ peaks of MoSe$_2$ overlaps those due to ZnSe and ZnS secondary phases,[22,23] respectively, making it difficult to identify the secondary phases in the film. By choosing an appropriate excitation energy matching resonance conditions, one can selectively enhance the signal from MoSe$_2$ and unambiguously identify the existence of MoSe$_2$ in Cu$_2$ZnSn(S,Se)$_4$.

Here we present Raman studies on bulk MoSe$_2$ using eight different excitation energies in the range of 1.6 – 3.8 eV. We found that the Raman spectrum changes dramatically at excitation energies near resonance with exciton states[24] because different modes are enhanced at different resonances. The results would provide a foundation on Raman studies of few-layer MoSe$_2$.

## Results & discussion

The crystal structure of 2H-MoSe$_2$ is shown in Fig. 1. Consecutive layers are stacked such that the Mo and Se atoms form a hexagonal arrangement in the top view as in Fig. 1(a). The point group of bulk 2H-MoSe$_2$ is $D_{6h}^4$, and there are 12 vibration modes at the center of the Brillouin zone, expressed as[19,25]

$$\Gamma = A_{1g} + 2A_{2u} + B_{1u} + 2B_{2g} + E_{1g} + 2E_{1u} + 2E_{2g} + E_{2u}. \tag{1}$$

$A_{2u}^1$ and $E_{1u}^1$ are acoustic modes, $A_{2u}^2$ and $E_{1u}^2$ are infrared active modes, $A_{1g}$, $E_{1g}$, $E_{2g}^1$, and $E_{2g}^2$ are Raman active modes, and the rest are inactive.[1] The Raman tensors[19] are expressed with respect to X=[100], Y=[010], and Z=[001] as

$$A_{1g} = \begin{pmatrix} a & 0 & 0 \\ 0 & a & 0 \\ 0 & 0 & b \end{pmatrix}, \quad E_{1g} = \begin{pmatrix} 0 & 0 & 0 \\ 0 & 0 & c \\ 0 & c & 0 \end{pmatrix}, \begin{pmatrix} 0 & 0 & -c \\ 0 & 0 & 0 \\ -c & 0 & 0 \end{pmatrix}, \qquad (2)$$

$$\text{and } E_{2g} = \begin{pmatrix} 0 & d & 0 \\ d & 0 & 0 \\ 0 & 0 & 0 \end{pmatrix}, \begin{pmatrix} d & 0 & 0 \\ 0 & -d & 0 \\ 0 & 0 & 0 \end{pmatrix}. \qquad (3)$$

In the backscattering geometry, $E_{1g}$ modes are forbidden.

Figure 2 shows the Raman spectra for different excitation energies, normalize to the intensity of the peak at 242 cm$^{-1}$, which is present in all the spectra. The most prominent peaks are the Raman active $A_{1g}$ and $E_{2g}^1$ modes[19] at 242 and 285 cm$^{-1}$, respectively. It is evident that these two modes have different excitation energy dependences as seen in Fig. 3(a). In order to further compare the excitation dependences of these two peaks, the intensities of the two peaks as well as their ratio are plotted as functions of the excitation energy in Fig. 3(b). In order to correct for the throughput of the measurement system, the intensities for different excitation energies were calibrated with the 520 cm$^{-1}$ signal from the Si substrate accounting for the resonance Raman curve of Si.[26]

The $A_{1g}$ peak is strong at the lowest and the highest excitation energies of 1.58 and 3.82 eV, respectively, with the minimum at 2.71 eV. On the other hand, the $E_{2g}^1$ peak is almost invisible at low excitation energies and increases monotonically at excitation energies above 2.54 eV. The excitation energy dependence of the Raman peak intensity is analyzed in terms of the resonance Raman effect. For bulk MoSe$_2$, the indirect gap is located close to 1.1 eV whereas the lowest direct transition is around 1.5 eV. For optical processes, however, since the excitonic effect significantly modifies the optical spectrum, the exciton states should be considered.[27] Optical transmission measurements estimated the energies of the *A*, *B*, and *C* exciton states of bulk MoSe$_2$ at 1.6, 1.8, and 2.8 eV, respectively.[24,28] Therefore, the excitation energy dependence of the $A_{1g}$ peak can be understood in terms of the ordinary resonance effect. The relative enhancement of the $E_{2g}^1$ peak for the excitation energies of 2.71 and 2.81 eV, however, calls for a special attention. Recently, Carvalho *et al.*[18] reported a similar effect in resonance Raman studies of MoS$_2$. They observed that unlike the $A_{1g}$ peak, the $E_{2g}^1$ peak exhibits a strong enhancement at the energy of the *C* exciton, which is associated with the transition between the top of the valence band and the first three lowest conduction bands near the Γ point. This peculiar difference between $A_{1g}$ and $E_{2g}^1$ modes was interpreted in terms of the symmetry of the atomic orbitals contributing to the exciton states and symmetry-sensitive exciton-phonon coupling. In MoSe$_2$, on the other hand, Beal *et al.*[24] interpreted a sharp feature at ~2.84 eV in the optical transmission spectrum as a band-to-band transition between higher energy bands. However, it is more reasonable to interpret this transition at ~2.84 eV as being due to the *C* exciton in MoSe$_2$ because the peculiar resonance behavior of the $E_{2g}^1$ mode and the lack thereof for the $A_{1g}$ mode are very similar to what was observed in MoS$_2$. Although there are differences in the band structures of MoS$_2$ and MoSe$_2$, the general features of the *C* exciton should be fairly similar and so the observed peculiar resonance of the $E_{2g}^1$ mode can be interpreted similarly.

In addition to the main peaks, several fine features are clearly resolved in the frequency range from 100 to 650 cm$^{-1}$. The measured peak positions are compared with theoretical calculations by Ding and Xiao[29] to assign the phonon modes. The assignments are summarized in Table 1.

The peaks at 169 cm$^{-1}$ is assigned to the $E_{1g}$ mode. This mode is forbidden in backscattering but appears due to the resonance effect. Since the $E_{1g}$ mode is an in-plane vibration mode similar to the $E_{2g}^1$ mode, its coupling with the *C* exciton state would be strongly enhanced, leading to the selection rule breaking near the resonance with the *C* exciton state. The selective enhancement of the infrared active and Raman forbidden $A_{2u}^2$ peak at 353 cm$^{-1}$ cannot be explained in the same manner. This peak was measured in infrared measurements[19] as well as in a recent Raman measurement with the 442-nm exciation.[5] When the excitation energy is close to the exciton energy, strongly localized exciton wavefunction effectively breaks the symmetry of the system, which in turn activates the modes that are normally Raman inactive. Similar symmetry breaking due to resonance with a strongly localized state was observed in nitrogen doped GaAs.[30] Therefore, we interpret it's the enhancement of this peak as being due to the resonance-induced symmetry breaking.

We also attempt to analyze weaker features, labelled *a*, *b*, *c*, and *d*. Because the peak positions do not match any possible first order Raman processes, these features are interpreted as being due to second order Raman processes. Its appearance correlates with that of the multiple phonon peaks *d* and similar peaks at higher energies discussed below. Its frequency is close to that of Stokes-anti-Stokes pair of $E_{2g}$ – LA at the M point. Similar features were observed in Raman scattering of MoS$_2$ as well.[17] The peak *b* at 238 cm$^{-1}$ appears as a shoulder for the 2.54 and 2.81 eV excitations and is well resolved for the 2.71 eV excitation. For all the other excitation energies, it is not resolved. The frequency difference between this peak and the $A_{1g}$ mode is 3 cm$^{-1}$. Sekine *et al.*[19] calculated the Davydov splitting of MoSe$_2$ using a simple spring model with two layers per unit cell and found its value to be ~2 cm$^{-1}$, which is in line with more recent calculations.[29] Therefore, we interpret this peak as the $B_{1u}$ mode. Tonndorf *et al.*[21] observed similar splitting in Raman spectra of 3-5 layer MoSe$_2$, measured with the 2.41-eV excitation. In addition to the $A_{1g}$ mode, the peak *c* at 248 cm$^{-1}$ is observed for all excitation energies. This peak has not been analyzed in previous studies. The positon of this peak is close to the calculated two-phonon energy of the $E_{2g}$ shear mode at the M point (248 cm$^{-1}$). For the 1.58-eV excitation, a broad peak labelled *d* and similar peaks at higher frequencies are observed. These peaks are multiple-phonon scattering peaks involving zone boundary phonons. For example, the frequency of the peak $d$, 315.3 cm$^{-1}$, is close to the sum of the $E_{1g}$ (181 cm$^{-1}$) branch and the LA (136 cm$^{-1}$) branch or the 2-phonon frequency of the $B_{2g}$ (314 cm$^{-1}$) branch at the M point. Similar multiple phonon peaks appear for the 2.71 and 2.81 eV excitations. In MoS$_2$, similar peaks due to 2-phonon scattering appear near resonances,[17] and we interpret these peaks also as being due to 2-phonon scattering activated by the resonance effect.

In summary, we have shown that the Raman spectrum of MoSe$_2$ varies greatly with the excitation energy. The enhancements of the Raman peaks are explained in terms of the resonance with various optical transitions including different types of exciton states. Different symmetries of the exciton states and the exciton-phonon coupling can explain selective enhancement of particular phonon modes for some excitation energies. Moreover, based on our results, we propose to use excitation energies in the range of 2.54 – 2.81 eV to take advantage of the resonant enhancement of the $E_{1g}$ or $E_{2g}^1$[1] peaks for detection of MoSe$_2$ secondary phases in thin film solar cell materials.

**Methods**

A single-crystal 2H-type bulk MoSe$_2$ flake bought from HQ Graphene was measured by micro-Raman spectroscopy. For the Raman measurement, we used 8 different excitation sources: the 325- and 441.6-nm (3.82 and 2.81 eV) lines of a He-Cd laser; the 457.9-, 488-, and 514.5-nm (2.71, 2.54, and 2.41 eV) lines of an Ar$^+$ laser; the 532-nm (2.33 eV) line of a diode-pumped-solid-state laser; the 632.8-nm (1.96 eV) line of a He-Ne laser; and the 784.8-nm (1.58 eV) line of a diode laser. The laser beam was focused onto the flake by using a 50× objective lens (0.8 N.A.) for all excitation wavelengths except for the 325-nm excitation for which a 40× uv objective lens (0.5 N.A.) was used. The scattered signal was collected by the same objective lens (backscattering geometry) and was dispersed with a Jobin-Yvon Horiba iHR550 spectrometer. Either a 1200-grooves/mm (630-nm blaze) or a 2400-grooves/mm (400-nm blaze) grating was chosen to acquire the best signal to noise ratio. A liquid-nitrogen-cooled back-illuminated charge-coupled-device detector was used, and thin-film interference filters (RazorEdge Filters) from Semrock were used to reject the Rayleigh-scattered light. The laser power was kept at 100 µW for all the measurement to avoid local heating of the sample. The spectral resolution was below 1 cm$^{-1}$.

**Figure 1.** (a) Top view and (b) side view of 2H-MoSe$_2$.

**Figure 2.** Raman spectra of a bulk MoSe$_2$ with different excitation energies. The intensity was normalized by the intensity of $A_{1g}$ mode of each spectrum. The sharp peaks below 150 cm$^{-1}$ for the 2.71-eV excitation are due to molecular vibration modes of air.

**Figure 3.** (a) Raman spectra of bulk MoSe$_2$ taken with the excitation energies indicated. (b) Intensities of $A_{1g}$, and $E_{2g}^1$ modes and $E_{2g}^1/A_{1g}$, ratio as functions of the excitation energy. Some major optical transition energies[24] are indicated by capital letters.

**Table 1.** Comparison of calculated vibrational modes with Raman peak positions in cm$^{-1}$ for different excitation energies.

| $E_{exc}$ (eV) | a $E_{2g}^1$ – LA at M | $E_{1g}$ | b $B_{1u}$ | $A_{1g}$ | c $2E_{2g}^2$ at M | $E_{2g}^1$ | d | $A_{2u}^2$ |
|---|---|---|---|---|---|---|---|---|
| cal$^{29}$ | 281-136=145 | 170 | 240 | 242 | 2×124=248 | 284.5 | 181+136=317$^a$ <br> 2×157=314$^b$ | 344 |
| 1.58 | 139.9 | 167.6 | - | 242.0 | 249.3 | 285.0 | 313.8 | - |
| 1.96 | 139.5 | 167.3 | - | 242.0 | 249.3 | 285.4 | 315.0 | - |
| 2.33 | - | 168.0 | - | 242.0 | 249.2 | 284.8 | 317.1 | - |
| 2.41 | - | 168.8 | - | 242.0 | 249.4 | 284.4 | - | 352.8 |
| 2.54 | - | 168.8 | - | 242.0 | 249.4 | 284.4 | - | 352.2 |
| 2.71 | 146.5 | 168.9 | 238.6 | 242.0 | 249.4 | 284.2 | - | 352.7 |
| 2.81 | 144.3 | 168.2 | - | 241.9 | 249.3 | 285.5 | - | 352.9 |
| 3.82 | - | - | - | 241.7 | not resolved | 284.7 | - | 351.9 |

$^a$ Sum of $E_{1g}$ and LA branches at the M point.
$^b$ 2-phonon frequency of the $B_{2g}$ branch at the M point.

## Acknowledgments

This work was supported by the New & Renewable Energy of the Korea Institute of Energy Technology Evaluation and Planning (KETEP) grant funded by the Korea government Ministry of Trade, Industry and Energy (No. 20123010010130) and the National Research Foundation (NRF) grant funded by the Korean government (MSIP) (No. 2011-0017605).

## Author contributions

D. N. and J.-U. L conducted the measurements. D.N., J.-U. L. and H. C. analyzed the data and discussed the results. All authors wrote the manuscript together.

## Additional Information

The authors declare no competing financial interests.

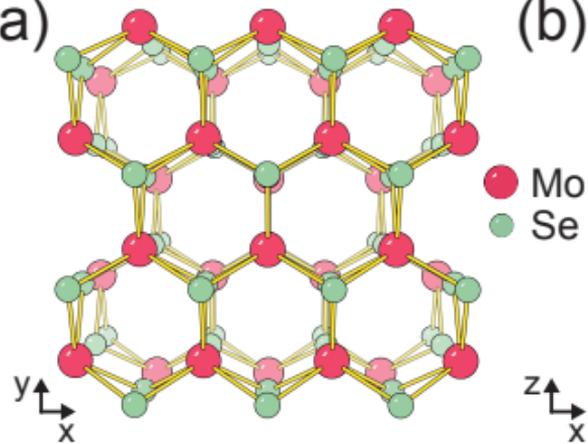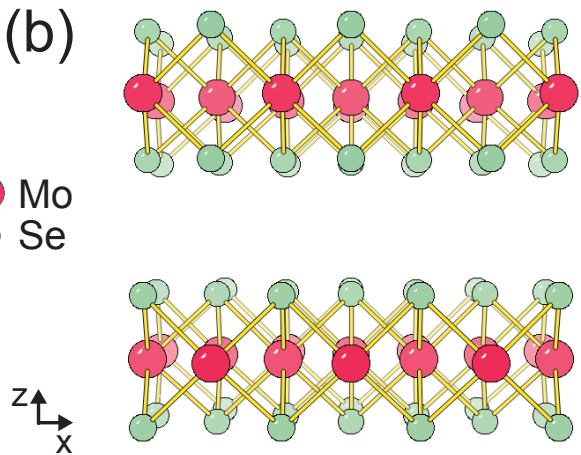

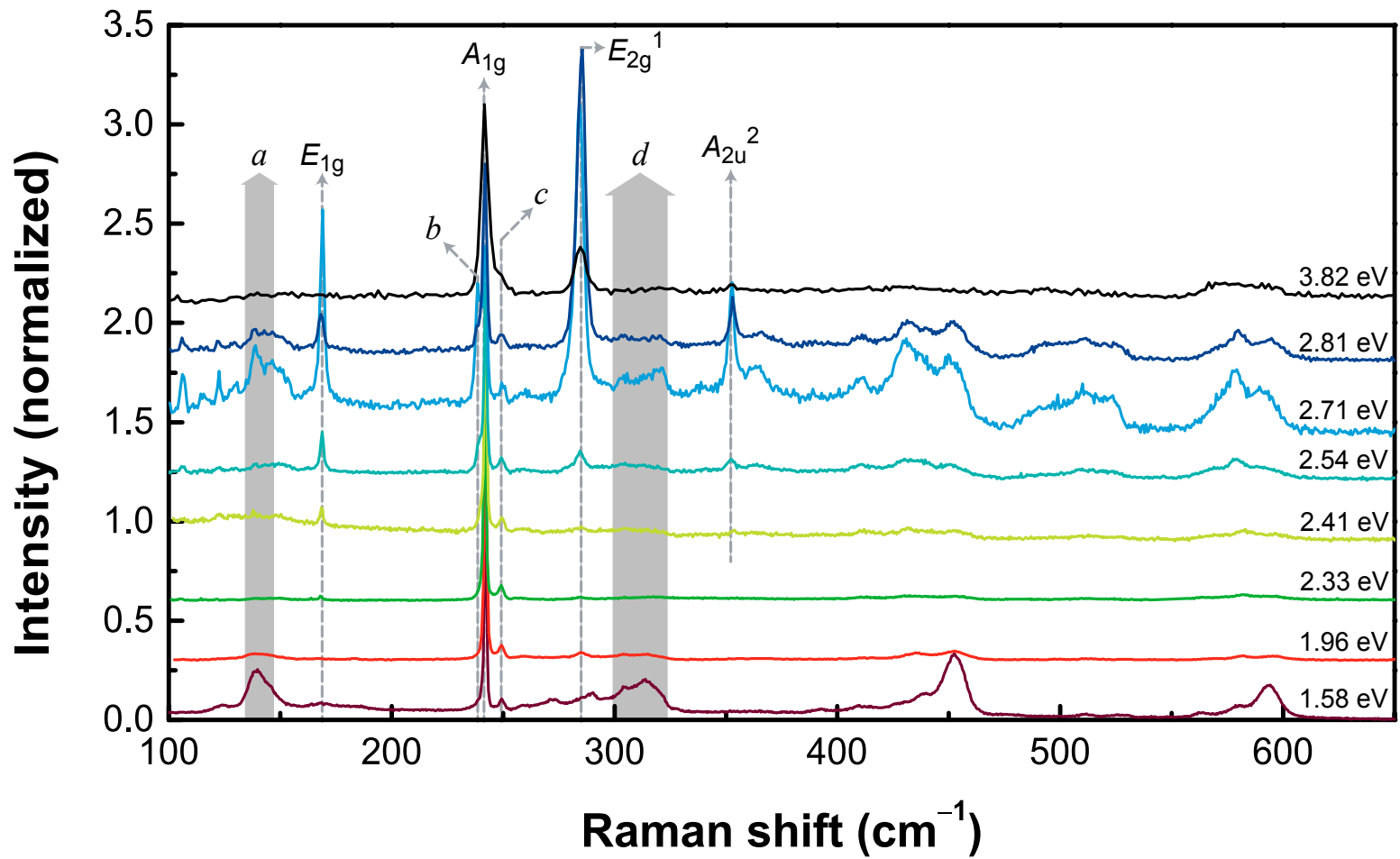

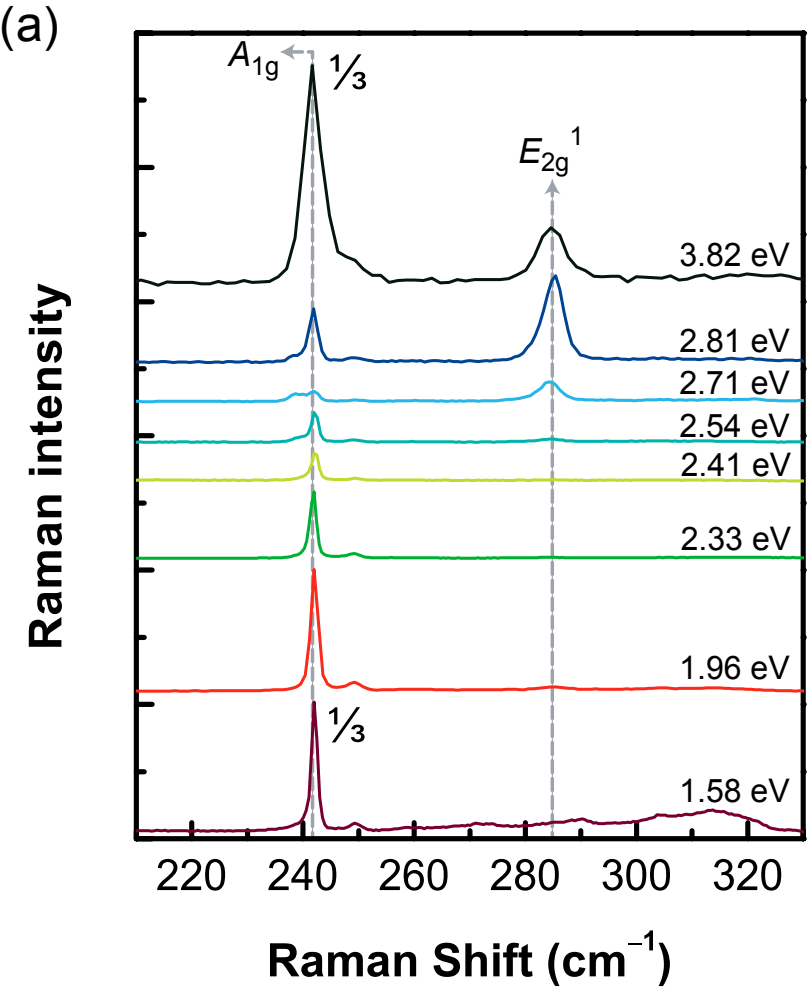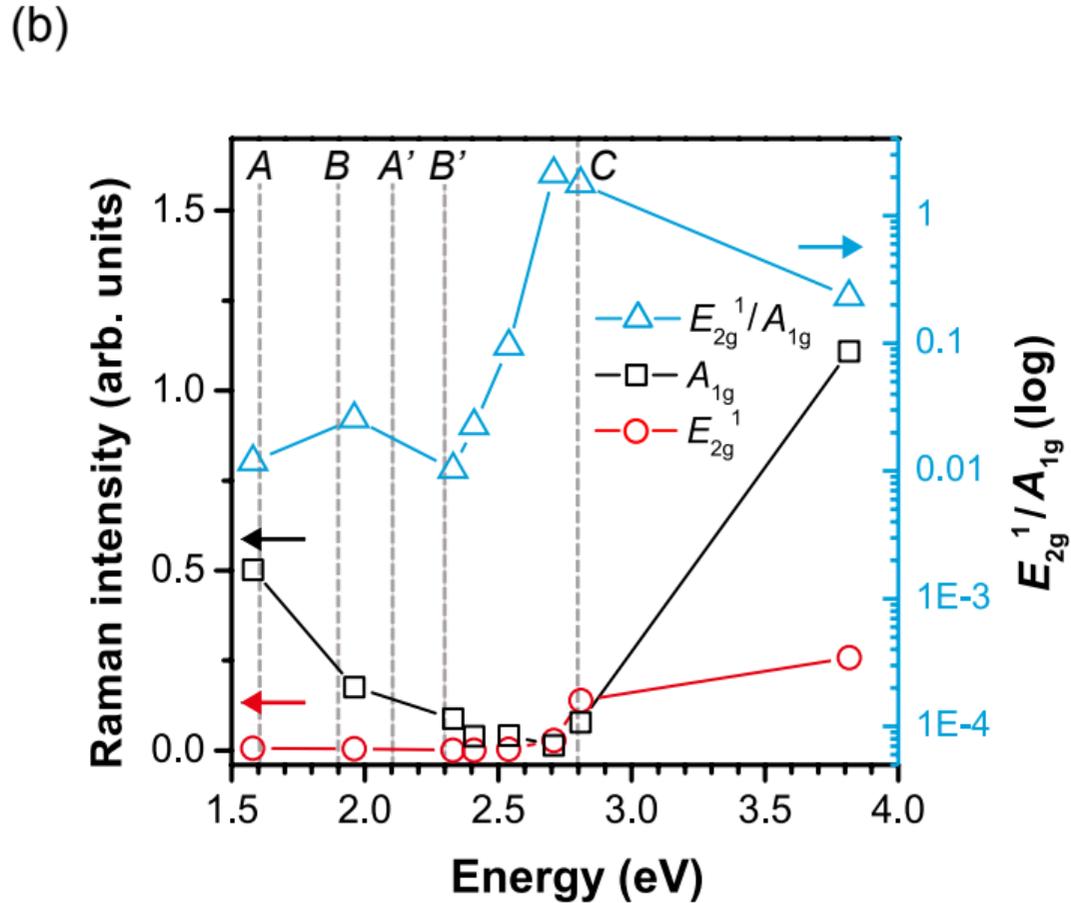